\documentclass[aps,showpacs,amsmath,amssymb,twocolumn,floatfix]{revtex4}
\usepackage{psfig,epsfig}
\newcommand{\ga}{\alpha}
\newcommand{\gb}{\beta}
\newcommand{\gc}{\gamma}
\newcommand{\gd}{\delta}
\newcommand{\gk}{\kappa}

\newcommand{\gl}{\lambda}
\newcommand{\gs}{\sigma}

\newcommand{\hh}{\frac{1}{2}}
\newcommand{\vk}[1]{\overline{#1}}
\newcommand{\uk}[1]{\underline{#1}}

\begin{document}
%\draft
\title{Light front thermal field theory at finite temperature
  and density} \author{M. Beyer, S. Mattiello}
\affiliation{Fachbereich Physik, Universit\"at Rostock, 18051 Rostock,
  Germany} \author{T. Frederico} \affiliation{Dep. de F\'\i sica,
  Instituto Tecnol\'ogico de Aeron\'autica,
  Centro T\'ecnico Aeroespacial, \\
  12.228-900 S\~ao Jos\'e dos Campos, S\~ao Paulo, Brazil.}
\author{H.J. Weber} \affiliation{ Dept. of Physics, University of
  Virginia, Charlottesville VA, U.S.A.}
\begin{abstract}
  We investigate quark matter at finite temperature and finite
  chemical potential as an example for a relativistic many-particle
  quantum system. Special relativity is realized through the front
  form that allows for a Hamiltonian formulation of a statistical
  operator. Utilizing our previous results we generalize the present
  formulation of a relativistic thermal field theory to include a
  finite chemical potential.  As an application we use the
  Nambu-Jona-Lasinio model to investigate the gap equation and chiral
  restoration.
  
\end{abstract}
\maketitle

\section{Introduction}
The front form~\cite{Dirac:cp} is widely recognized as an alternative
form of a relativistic theory. It has been successfully applied in
particular in the context of quantum chromo dynamics (QCD), e.g., to
describe deep inelastic scattering or to model the structure of
hadrons, see e.g. Ref.~\cite{Brodsky:1997de}.  The front form provides
a rigorous relativistic many-body theory that has been applied, e.g.,
to investigate nuclear matter~\cite{Miller:1999ap,Miller:2000kv}.
Therefore it is quite natural to ask for the merits of light-front
quantization in the context of quantum statistics to describe
many-particle systems of finite temperature $T$ and finite chemical
potential $\mu$.  There is a need to do so, in particular, to model
the evolution of the early universe, the dynamics of relativistic
heavy ion collisions, multi-particle correlations in such a collision,
relativistic Coulomb plasma, among others.

The difficulty to generalize the statistical operator to the light-front is
related to the longitudinal mode $k^+=k^0+k^3$. Whereas temperature for the
transverse modes related to $\vec k_\perp=(k^1,k^2)$ can easily be defined,
the longitudinal modes, can only by related to the instant form by a limiting
process (infinite momentum frame).  Therefore the longitudinal modes appear in
a proper definition of the statistical operator on the light-front. As a
consequence the Fermi function (here given for a grand canonical ensemble)
explicitly depends on the light-front energy $k^-_{\mathrm{on}}$,
\begin{equation}
k^-_{\mathrm{on}}=\frac{\vec k_\perp^2+m^2}{k^+}
\label{eqn:kon}
\end{equation}
and the light-front ``mass'' $k^+$, viz.~\cite{Beyer:2001bc}
\begin{equation}
f(k^+,\vec k_\perp)=
\left[\exp\left\{\frac{1}{T}\left(\frac{1}{2}k^-_{\mathrm{on}}
+\frac{1}{2}k^+-\mu\right)\right\}+1\right]^{-1}.
\label{eqn:fermi}
\end{equation}
This function is consistent with the instant form and has already been used to
investigate the formation of three-quark clusters (nucleons) in the hot and
dense phase of nuclear matter close to the QCD phase
transition~\cite{Beyer:2001bc,Mattiello:2001vq}.

Recently, Brodsky~\cite{Brodsky:2001ww,Brodsky:2002} suggested a partition
function for a canonical ensemble ($\mu=0$) that would lead to a different
Fermi function (naive generalization). It has been explicitly shown that the
naive generalization has serious problems, because the Fermi function leads to
divergent integrals~\cite{Alves:2002tx}, which is not the case for the
distribution function given in Eq.~(\ref{eqn:fermi}) due to the presence of
the $k^+$ term. Presently, there is a renewed interest and discussion on how
to formulate thermal field theory on the
light-front~\cite{Weldon:2003uz,Weldon:2003vh,Das:2003mf,Kvinikhidze:2003wc}.
So far, none of these generalizations consider finite chemical potentials that
are relevant for QCD.  Utilizing our previous results we generalize the
present formulation of thermal field theory on the light-front to include a
finite chemical potential.

As an example, we use the Nambu-Jona-Lasinio (NJL)
model~\cite{Nambu:tp,Nambu:fr} that is a powerful tool to investigate
the non-perturbative region of QCD as it exhibits spontaneous breaking
of chiral symmetry and the appearance of Goldstone bosons in a
transparent way.  The region of finite chemical potential and
temperature can be investigated in relativistic heavy ion collisions
of the newly constructed (RHIC) and upcoming machines (LHC, GSI). It
is particularly interesting since the phase structure is expected to
be
complex~\cite{Alford:1997zt,Alford:1998mk,Rapp:1997zu,Rajagopal:2000wf}.
It might be also relevant, e.g. for the physics of compact stars.
Lattice QCD studies of this region ($\mu\neq 0$) are far more limited
than for $\mu=0$~\cite{Karsch:1999vy}, although some progress has been
achieved recently, see e.g.~\cite{Fodor:2002sd} and refs. therein.

In Sec.~\ref{sec:statphys} we use our previous results to generalize
the present formalism of thermal field theory on the light-front to
finite chemical potentials. In Sec.~\ref{sec:model} we briefly
introduce the NJL model on the light-front to give an example. This
will be done along the lines of Ref.~\cite{Bentz:1999gx} and in
Sec.~\ref{sec:results} we will present numerical results for the quark
mass at finite temperature and densities (chiral restoration).

\section{Light-front statistical physics}
\label{sec:statphys}

Here we present more details of our previous results, and provide the derivation of
the distribution function for the grand canonical ensemble.  See the Appendix
~\ref{app:notation} for notation.  The four-momentum operator $P^\mu$ on the
light-front  is given by
\begin{equation}
P^\mu=\int dS_+\; T^{+\mu}(x),
\label{eqn:P}
\end{equation}
where $T^{\mu\nu}(x)$ denotes the energy momentum tensor defined
through the Lagrangian of the system and $S_\mu$ is the quantization
surface, see Eq.~(\ref{eqn:dS}).  Hence the Hamiltonian is given by
\begin{equation}
P^-
=\int dS_+\; T^{+-}(x).
\end{equation}
To investigate a grand canonical ensemble it is necessary to define
the number operator. The number operator on the light-front is given by
\begin{equation}
N=\int dS_+\; J^+(x),
\label{eqn:N}
\end{equation}
where $J^\nu(x)$ is the conserved current. These are the necessary
ingredients to generalize the covariant calculations at finite
temperature~\cite{Israel:1976tn,isr81,Weldon:aq} to the light-front.
The grand canonical partition operator on the
light-front is given by
\begin{equation}
Z_G = \exp\left\{\int dS_+\;
[- \gb_\nu T^{+\nu}(x)+\ga J^+(x) ]\right\},
\end{equation}
where $\ga=\mu/T$, with the chemical potential $\mu$. The velocity of the medium is
\begin{equation}
\gb_\nu = \frac{1}{T} u_\nu
\end{equation}
with the normalized time-like vector $u_\nu u^\nu
=1$~\cite{Israel:1976tn}. We choose   
\begin{equation}
u^\nu = (u^-,u^+,\vec u^{\perp})= (1,1,0,0),
\end{equation}
hence $u^+-u^-=2u^3=0$, i.e. the medium is at rest. The grand partition
operator becomes
\begin{equation}
Z_G =  e^{-K/T},\quad K\equiv\frac{1}{2}(P^-+P^+) - \mu N
\label{eqn:ZG}
\end{equation}
with $P^\pm$ and $N$ defined in Eqs.~(\ref{eqn:P}) and (\ref{eqn:N}).  To
further evaluate Eq.~(\ref{eqn:ZG}), the operators will be used in Fock space
representation.  For an ideal gas of Fermions we use the solution of the free
Dirac field (conventions of~\cite{Lepage:1980fj})
\begin{eqnarray}
\Psi_\ga(x)&=&\sum_\lambda \int \frac{dk^+d^2k_\perp}{\sqrt{2k^+(2\pi)^3}}
\Big(b_\lambda(\vk{k})
u_{\ga \lambda}(\vk{k})e^{-ik_\mathrm{on}\cdot x}
\nonumber\\&&\qquad\qquad\qquad
+d^\dagger_\lambda(\vk{k})
v_{\ga \lambda}(\vk{k})e^{ik_\mathrm{on}\cdot x}
\Big)
\end{eqnarray}
with the four-vector $k_\mathrm{on}=(k^-_\mathrm{on},\vk{k})$, the
abbreviation $\vk{k}=(k^+,\vec k_\perp)$, and the Fock operators
\begin{eqnarray}
  \{b_\lambda(\vk{k}),b^\dagger_\gs(\vk{p})\}&=&
\gd(k^+-p^+)\gd^{(2)}(\vec k_\perp-\vec p_\perp)
\gd_{\lambda\sigma},\\
  \{d_\lambda(\vk{k}),d^\dagger_\gs(\vk{p})\}&=&
\gd(k^+-p^+)\gd^{(2)}(\vec k_\perp-\vec p_\perp)
\gd_{\lambda\sigma}.
\end{eqnarray}
 The Hamiltonian takes the form
\begin{equation}
P^-=\sum_\lambda \int dk^+d^2k_\perp\; k^-_\mathrm{on}
\left(b^\dagger_\lambda(\vk{k})
b_\lambda(\vk{k})
+d^\dagger_\lambda(\vk{k})
d_\lambda(\vk{k})\right)
\end{equation}
and for the plus component of the Dirac current,
\begin{equation}
N=\sum_\lambda \int  dk^+d^2k_\perp\; 
\left(b^\dagger_\lambda(\vk{k})b_\lambda(\vk{k})
-d^\dagger_\lambda(\vk{k})d_\lambda(\vk{k})\right).
\end{equation}
Inserting all the above into (\ref{eqn:ZG}) gives
\begin{eqnarray}
K&=&\sum_\lambda\int dk^+d^2k_\perp\;
\Big(\gk^+\;b^\dagger_\lambda(\vk{k})b_\lambda(\vk{k})
\nonumber\\&&\qquad\qquad\qquad
+\gk^-\;d^\dagger_\lambda(\vk{k})d_\lambda(\vk{k})\Big),
\\
\gk^\pm&=&\frac{1}{2}(k^-_\mathrm{on}+k^+) \mp \mu.
\end{eqnarray}
%% \begin{eqnarray}
%% K&=&\sum_\lambda\int dk^+d^2k_\perp\;
%% \left\{\left(\frac{1}{2}(k^-_\mathrm{on}+k^+) - \mu \right)
%% \gk^+\;b^\dagger_\lambda(\vk{k})b_\lambda(\vk{k})
%% \right.\nonumber\\&&\qquad\qquad\left.+
%% \left(\frac{1}{2}(k^-_\mathrm{on}+k^+) + \mu \right)
%% \gk^-\;d^\dagger_\lambda(\vk{k})d_\lambda(\vk{k})
%% \right\},
%% \end{eqnarray}
The density operator for a grand canonical
ensemble~\cite{kad62,fet71}  in equilibrium
follows
\begin{equation}
\rho_G=(\mathrm{Tr}\ e^{-K/T})^{-1}\;e^{-K/T}.
\label{eqn:grand}
\end{equation}

The  light-cone time-ordered Green function for fermions is
\begin{equation}
{\cal G}_{\ga\gb}(x-y)=
\theta(x^+-y^+)\;{\cal G}^>_{\ga\gb}(x-y)
+ \theta(y^{+}-x^{+})\;{\cal G}^<_{\ga\gb}(x-y),
\label{eqn:defCrono}
\end{equation}
where
\begin{eqnarray}
{\cal G}^>_{\ga\gb}(x-y)&=&
i\langle\Psi_\ga(x)\bar\Psi_\gb(y)\rangle
\nonumber\\
{\cal G}^<_{\ga\gb}(x-y)&=&
-i\langle\bar\Psi_\gb(y)\Psi_\ga(x)\rangle,
\label{eqn:defG}
\end{eqnarray}
We note here that the light-cone time-ordered Green function differs from the
Feynman propagator $S_F$ in the front form by a contact term (which is not the
case in the instant form).  Evaluating Eq.~(\ref{eqn:defG}) for the vacuum, i.e.
$\langle\dots\rangle=\langle0|\dots|0\rangle$ (isolated case), the
 Green function is
\begin{widetext}
\begin{eqnarray}
i{\cal G}(x-y)&=
&\int \frac{d^4k}{(2\pi)^4} e^{-ik\cdot(x-y)}
\left(
\frac{\gamma k_\mathrm{on}+m}{\hh k^--\hh k^-_\mathrm{on}+i\varepsilon}
\frac{\theta(k^+)}{2k^+}
+\frac{\gamma k_\mathrm{on}+m}{\hh k^--\hh k^-_\mathrm{on}-i\varepsilon} 
\frac{\theta(-k^+)}{2k^+}
\right)
\nonumber\\&=&
\int \frac{d^4k}{(2\pi)^4} e^{-ik\cdot(x-y)}
\left(S_F(k)-\frac{\gamma^+}{2k^+}\right)\equiv
\int \frac{d^4k}{(2\pi)^4} e^{-ik\cdot(x-y)}
\;G(k^-,\vk{k})
\label{eqn:lfprop}
\end{eqnarray}
 \end{widetext}
where the Feynman propagator is given by
\begin{equation}
S_F(k)=\frac{\gamma k+m}{k^2-m^2+i\epsilon}.
\end{equation}
Eq.~(\ref{eqn:lfprop}) coincides with the light-front propagator given
previously in Ref.~\cite{chang}. Using the independent components $\Lambda^+
\Psi$ instead of $\Psi(x)$ in the definition of the chronological Green
function Eqs.~(\ref{eqn:defCrono}) and (\ref{eqn:defG}) leads to a different
propagator suggested, e.g.,  in~\cite{Srivastava:2000cf,Alves:2002tx}.

To evaluate the ensemble average
$\langle\dots\rangle=\mathrm{Tr}(\rho_G\dots)$ of
Eqs.~(\ref{eqn:defCrono}) and (\ref{eqn:defG}) (in-medium case), we
utilize the imaginary time formalism~\cite{kad62,fet71}. We
rotate the light-front time of the Green function to imaginary values,
i.e. $u\cdot x\rightarrow i\tau$. The anti-periodic boundary condition of
the imaginary time Green function holds on the light-front as well.
Hence the $k^-$-integral is replaced by a sum of light-front Matsubara
frequencies $\omega_n$ according to~\cite{Beyer:2001bc},
\begin{eqnarray}
  \frac{1}{2} k^- \rightarrow i\omega_n - \frac{1}{2} k^+ +\mu\equiv \frac{1}{2} k_n^-,
\end{eqnarray}
where $\omega_n=\pi \lambda T$, $\lambda=2n+1$ for fermions [$\lambda=2n$ for
bosons].
For noninteracting Dirac fields
the imaginary time Green function becomes
\begin{widetext}
\begin{eqnarray}
G(k^-_n,\vk{k})&=&
\frac{\gamma k_\mathrm{on}+m}{\hh k^-_n-\hh k^-_\mathrm{on}+i\varepsilon}
 \frac{\theta(k^+)}{2k^+}
(1-f^+(\vk{k}))
+\frac{\gamma k_\mathrm{on}+m}{\hh k^-_n-\hh k^-_\mathrm{on}-i\varepsilon} 
\frac{\theta(k^+)}{2k^+}
f^+(\vk{k})
\label{eqn:lfmed}\\
&&
+\frac{\gamma k_\mathrm{on}+m}{\hh k^-_n-\hh k^-_\mathrm{on}+i\varepsilon} 
\frac{\theta(-k^+)}{2k^+}
f^-(-\vk{k})
+\frac{\gamma k_\mathrm{on}+m}{\hh k^-_n-\hh k^-_\mathrm{on}-i\varepsilon} 
\frac{\theta(-k^+)}{2k^+}
(1-f^-(-\vk{k})).
\nonumber
\end{eqnarray}
\end{widetext}
 For a grand canonical ensemble the Fermi distribution
functions of particles $f^+\equiv f$ and antiparticles $f^-$ are given by
\begin{eqnarray}
f^\pm(k^+,\vec k_\perp)&=&\left[e^{\gk^\pm/T}+1\right]^{-1}
\nonumber\\&=&
\left[\exp\left\{\frac{1}{T}\left(\frac{1}{2}k^-_{\mathrm{on}}
+\frac{1}{2}k^+\mp\mu\right)\right\}+1\right]^{-1}
\label{eqn:fermipm}
\end{eqnarray}
and $k^-_\mathrm{on}$ of Eq.~(\ref{eqn:kon}). The (particle) fermionic
distribution function on the light-front has been given in
Eq.~(\ref{eqn:fermi}). The propagator for this case ($f^-=0$) has been given
previously and used to investigate the stability and dissociation of a
relativistic three-quark system in hot and dense quark
matter~\cite{Beyer:2001bc,Mattiello:2001vq}.  Note that for the canonical
ensemble, $\mu=0$, this result coincides with the one given more recently
in~\cite{Alves:2002tx} (up to different conventions in the metric).

For equilibrium the imaginary time formalism and the real time formalism are
linked by the spectral function~\cite{kad62,fet71}. 
  
\section{NJL model on the light-front}
\label{sec:model}

The Nambu-Jona-Lasinio (NJL) originally suggested
in~\cite{Nambu:tp,Nambu:fr} has been reviewed in
Ref.~\cite{klevansky:1992} as a model of quantum chromo dynamics (QCD),
where also a generalization to finite temperature and finite chemical
potential has been discussed. Most of the applications of the NJL
model have utilized the instant form quantization. On the other hand
light-front quantization of quantum field theories has emerged as a
promising method for solving problems in the strong coupling regime.
Examples and in particlar the light-front formulation of QCD, are
reviewed in~\cite{Brodsky:1997de}. Based on a Hamiltonian the
light-front approach is particularly suited to treat bound states
(correlations), the most relevant consisting of $q\bar q$ and $qqq$
valence quarks. The formation of bound states is in particular
interesting in the vicinity of the quark-hadron phase transition.  The
front form is useful in the context of quantum statistics, since the
Fock space representation allows a consistent formulation of
thermodynamic properties, as demonstrated in the previous Section.
Since the NJL model, on one hand, is a powerful tool to investigate
the non-perturbative region of QCD and, on the other hand, is
comparatively transparent and simple, we use it here as an example to
tackle the following questions~\cite{Beyer:2001bc}: How are hadrons
formed/dissociated? Is the formation/dissociation transition at the
same place in the phase diagram as the chiral phase transition?
Approaches with a closer connection to QCD than the NJL model, based,
e.g., on the method outlined in Ref.~\cite{Brodsky:1997de}, have to
follow. These may be of use as either complementary or supplementary
approaches to lattice QCD (e.g.  at large chemical potentials $\mu$,
where lattice QCD is still not applicable).

The NJL
Lagrangian is given by
\begin{equation}
  \label{eqn:NJL}
  {\cal L}=\bar\psi (i\gc \partial -m_0)\psi 
+ G\left( (\bar\psi\psi)^2 + (\bar\psi i\gc_5 \tau\psi)^2\right). 
\end{equation}
In mean field approximation the gap equation is
\begin{equation}
  \label{eqn:gap}
  m=m_0 - 2G\langle \bar\psi\psi\rangle
= m_0 + 2 iG\gl\int\frac{d^4k}{(2\pi)^4}\;\mathrm{Tr}S_F(k),
\end{equation}
where $\gl=N_f N_c$ in Hartree and $\gl=N_f N_c+{\textstyle\frac{1}{2}}$ in
Hartree-Fock approximation, $N_c$ ($N_f$) is the number of colors (flavors).

\subsection{Isolated case}
On the light-front the realization of chiral symmetry breaking is a subtle
issue~\cite{Dietmaier:hv,Itakura:1997zw,Itakura:2000te,Lenz:2000em}.  Unlike
the instant form, for a given momentum $(k^+,\vec k_\perp)$, $k^+\neq 0$, the
dispersion relation on the light-front (\ref{eqn:kon}) is unambiguously
determined.  This is usually denoted as the simple vacuum structure on the
light-front. A subtlety is related to the zero modes, $k^+=0$, which contain
the information on symmetry breaking. For a recent review on the realization
of chiral symmetry breaking on the light-front see, e.g.,
Ref.~\cite{Itakura:2001yt}. Here we basically follow Ref.~\cite{Bentz:1999gx}.

After performing the $k^-$ integration the gap equation becomes ($x>0$)
\begin{equation}
  \label{eqn:gaplf}
  m-\tilde m_0=
2\tilde G \gl \int 
\frac{dxd^2 k_\perp}{2x(2\pi)^3}
  \;4m,
\end{equation}
where in addition $m_0\rightarrow \tilde m_0$ and $G\rightarrow \tilde
G$ have been renormalized. The parameters $m_0,G$ and $\tilde
m_0,\tilde G$ are related via $1/N_c$ expansion (see appendix B
of~\cite{Bentz:1999gx}). The explicit relation is given below for
the regularization schemes discussed here. Note that due to the trace
either $G(k)$ or $S_F(k)$ of Eq.~(\ref{eqn:lfprop}) might be used,
since $\mathrm{Tr}\gc^+=0$. The integral of Eq.~(\ref{eqn:gaplf})
clearly diverges and one has to introduce a regularization scheme.
Several regularization schemes have been in use. A common scheme has
been suggested by Lepage and Brodsky~\cite{Lepage:1980fj} (LB) 
\begin{equation}
\vec k_\perp^2<\Lambda^2_\mathrm{LB} x(1-x)-m^2,
\label{eqn:LB}
\end{equation}
which implies $x_1\le x\le x_2$, where
\begin{equation}
x_{1,2}=\hh \left(1\mp\sqrt{1-4m^2/\Lambda^2_\mathrm{LB}}\right).
\label{eqn:x}
\end{equation}
The equivalence of the light front LB scheme to the instant form three
momentum (3M) scheme, where (\ref{eqn:gap}) is integrated over $k^0$ and $
\mathbf{k}^2<\Lambda^2_{\mathrm{3M}}$ has been shown in
Ref.~\cite{Bentz:1999gx} utilizing the Sawicki
transformation~\cite{sawicki:1985}.  To compare results note that for the
mentioned regularization schemes the following identifications hold
\begin{eqnarray}
\tilde m_{0,\mathrm{LB}}&=&m_{0,\mathrm{3M}},
\label{eqn:m03M}
\\
\tilde G_\mathrm{LB}&=&G_\mathrm{3M},\\
\Lambda^2_\mathrm{LB}&=&4(\Lambda^2_\mathrm{3M}+m^2).
\label{eqn:L3M}
\end{eqnarray}
Numerical values are
given in the next section. The calculation of
the pion mass $m_\pi$, the pion decay constant $f_\pi$, and the condensate
value follows~\cite{Bentz:1999gx}.

\subsection{In-medium case}
Here we are concerned in an application of the formalism to chiral
restoration in hot and dense matter.  The statistical approach shown
in the previous section for finite temperature $T$ and chemical
potential $\mu$ will be used.  Because of the medium, the propagator
to be used in the gap equation (\ref{eqn:gap}) is given in
Eq.~(\ref{eqn:lfmed}). The gap equation becomes
\begin{widetext}
\begin{equation}
  m(T,\mu)=\tilde m_0+2\tilde G\gl\int 
\frac{dk^+d^2k_\perp}{2k^+(2\pi)^3} \;4m(T,\mu)
%\nonumber\\&&\qquad\times
(1-f^+(k^+,\vec k_\perp)-f^-(k^+,\vec k_\perp)).
  \label{eqn:gapmed}
\end{equation}
\end{widetext}
The quark mass changes with the parameters $T$ and $\mu$. Note that being a
single particle distribution the Fermi functions $f^\pm$ that appear in the
gap equation depend on $k^+$, rather than $x=k^+/P^+$. Therefore the LB
regularization scheme cannot be used without modification that will be shown
below. To regularize Eq.~(\ref{eqn:gapmed}) we require instead
\begin{equation}
k^-_\mathrm{on}+ k^+<2\Omega.
\end{equation}
As a consequence $k^+_1<k^+<k^+_2$ and
\begin{eqnarray}
\vec k^2_\perp &<& 2\Omega k^+- (k^+)^2 -m^2,
\label{eqn:regperp}
\\
k^+_{1,2}&=&\Omega\mp\sqrt{\Omega^2-m^2}.
\label{eqn:reglong}
\end{eqnarray}
To connect this $\Omega$ scheme with the LB and the 3M schemes we use the Sawicki
transformation~\cite{sawicki:1985}. 
\begin{eqnarray}
x&=&\frac{k^+}{P^+}=\hh\left(1+\frac{k^3}{\omega_k}\right),\\
\omega_k&=&+\sqrt{\vec k_\perp^2+(k^3)^2+m^2},
\end{eqnarray}
where $P^+$ is evaluated in the rest system of the two particles involved in
the gap equation because of the loop, $P^+=2\omega_k$.  Starting from
Eq.~(\ref{eqn:gapmed}) with the regularization conditions (\ref{eqn:regperp})
and (\ref{eqn:reglong}) we substitute $k^+$ by $x$ and replace the regularization
conditions accordingly. If we choose 
\begin{equation}
\Lambda_\mathrm{LB}=2\Omega,
\end{equation}
we obtain the LB conditions (\ref{eqn:LB}) and (\ref{eqn:x}). The resulting
regularized gap equation in the LB scheme is
\begin{widetext}
\begin{eqnarray}
  m(T,\mu)&=&\tilde m_{0,\mathrm{LB}}+2\tilde G_\mathrm{LB}\gl\int\limits_\mathrm{LB}
\frac{dxd^2k_\perp}{2x(2\pi)^3} \;4m(T,\mu)\;
(1-f^+_\mathrm{LB}(x,\vec k_\perp)-
f^-_\mathrm{LB}(x,\vec k_\perp)).\\
f^\pm_\mathrm{LB}(x,\vec k_\perp)&=&
\left[\exp\left\{\frac{1}{T}\left(\frac{k^2_\perp+m^2}{2x\Lambda_\mathrm{LB}}
+\frac{x\Lambda_\mathrm{LB}}{2}\mp\mu\right)\right\}+1\right]^{-1},\\
\end{eqnarray}
In the LB regularized gap equation the Fermi functions depend explictly on the
regularization parameter $\Lambda_\mathrm{LB}$, because of the $k^+$
dependence of the Fermi function. Replacing  $k^+$ by $k^3$ in
Eq.~(\ref{eqn:gapmed}) we get
\begin{eqnarray}
  m(T,\mu)&=&m_{0,\mathrm{3M}}+2G_\mathrm{3M}\gl\int\limits_\mathrm{3M}
\frac{d^3k}{2\omega_k(2\pi)^3} \;4m(T,\mu)\;
(1-f^+(\omega_k)-f^-(\omega_k)),\label{eqn:gap3M}\\
f^\pm(\omega_k)&=&
\left[\exp\left\{\frac{1}{T}\left(\omega_k\mp\mu\right)\right\}+1\right]^{-1},
\end{eqnarray}
\end{widetext}
and using (\ref{eqn:regperp}) and (\ref{eqn:reglong}) the 
regularisation condition follows
\begin{equation}
\mathbf{k}^2<\Omega^2-m^2\equiv\Lambda_\mathrm{3M}^2.
\end{equation}
This regularized version (\ref{eqn:gap3M}) is analytically the same equation
as given previously in Ref.~\cite{klevansky:1992}. Therefore we have shown
explicitly that also in medium the $\Omega$ regularized light front gap
equation leads to the same results as the 3M regularized instant form gap
equation. We remark here also that, because of the $T$ and $\mu$ dependence of
the mass, the regulators $\Omega$ and $\Lambda_\mathrm{LB}$ need to depend
parametrically on temperature and chemical potential to keep the equivalence with
the 3M scheme.

\section{Results}
\label{sec:results}

\begin{table}[b]
\caption{\label{tab:parm} Parameters of the NJL model used in this
  analysis. $\Omega$ quoted for the isolated case.}
\begin{ruledtabular}
\begin{tabular}{ccc}
$\tilde G [10^{-6}$MeV]&$\tilde m_0$ [MeV]&$\Omega$ [MeV]\\\hline
5.51&5.67&714
\end{tabular}
\end{ruledtabular}
\end{table}

 \begin{figure}[tbp]
   \begin{center}
     \epsfig{figure=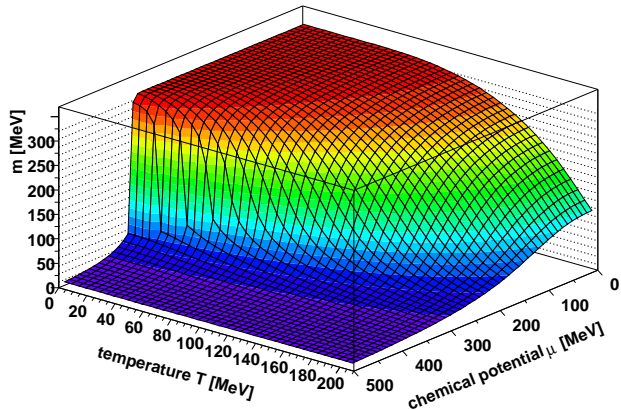, width=0.5\textwidth}
     \caption{\label{fig:mass} 
       Effective quark mass as a function of temperature and chemical
       potential. The fall-off is related to the vanishing condensate
       $\langle \bar u u\rangle$, which shows the onset of chiral
       symmetry restoration.  Critical temperature at $\mu=0$ is
       $T_c\simeq 190$ MeV.}
   \end{center}
 \end{figure}
\begin{figure}[tbp]
   \begin{center}
     \epsfig{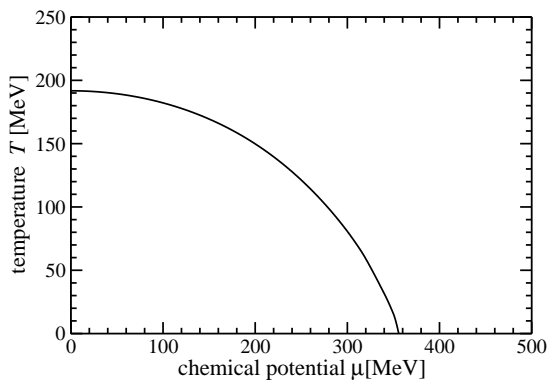}
     \caption{\label{fig:phase} 
       Chiral phase transition as defined in~\cite{asakawa:1989}. The lower
       part is the chiral broken phase, whereas the upper part reflects the
       restored phase.}
   \end{center}
 \end{figure}

The model parameters are adjusted to the isolated system. We use the Hartree
approximation, i.e. $\gl=N_cN_f=6$. Parameter values shown in
Table~\ref{tab:parm} are chosen to reproduce the pion mass $m_\pi=140$ MeV,
the decay constant $f_\pi=93$ MeV, and to give a constituent quark mass of
$m=336$ MeV. The difference between the bare mass $\tilde m_0$ and the
constituent mass is due to the finite condensate, which is $\langle \bar
uu\rangle^{1/3}=-247$ MeV for the parameters given in Table~\ref{tab:parm}.
The parameters are reasonably close to the cases used in the review
Ref.~\cite{klevansky:1992}.

In hot and dense quark matter the surrounding medium leads to a change of the
constituent quark mass due to the quasiparticle nature of the quark.  The
constituent mass as solution of Eq.~(\ref{eqn:gapmed}) is plotted in
Fig.~\ref{fig:mass} as a function of temperature and chemical potential. The
fall-off is related to chiral symmetry restoration, which would be complete
for $m_0=0$. It is related to the QCD phase transition. For $T\lesssim 60$ MeV
the phase transition is first order, which is reflected by the steep change of
the constituent mass. To keep close contact with the 3M results we have chosen
for the $\Omega$ in-medium regulator mass
$\Omega^2(T,\mu)=\Lambda_\mathrm{3M}^2+m^2(T,\mu)$ with
$\Lambda_\mathrm{3M}=630$ MeV fixed for all $T$ and $\mu$.

We define the phase transition to occur at a temperature at which $m(T,\mu)$
is half of the isolated constituent quark mass~\cite{asakawa:1989}. Because of
the above mentioned equivalence we do not expect any qualitative difference to
the results of the instant form analysis provided, e.g. by
Refs.~\cite{klevansky:1992,schwarz:1999}. The phase diagram is shown in
Fig.~\ref{fig:phase}. The line indicates the phase transition between the
chiral broken and the unbroken phase. 

\section{conclusion}
\label{sec:conclusion}
We have given a relativistic formulation of thermal field theory utilizing the
light front form. The proper partition operator (and the statistical operator)
have been given for the grand canonical ensemble. The special case of a
canonical ensemble is given for $\mu=0$. The resulting Fermi function depends
on transverse and also on the $k^+$ momentum components. The $k^+$ components
emerge in a natural way in a covariant approach and are also essential to
fulfill the light front analog of the Thouless criterion~\cite{thouless} for
the appearance of Cooper poles (color superconductivity)~\cite{prep}. As an
application we have revisited the NJL low energy model of QCD. Because of the
zero range interaction a regularization is necessary. The equivalence of
instant form 3M regularization and light front $\Omega$ regularization
utilized here for the gap equation holds for a $T$ and $\mu$ dependent cut-off
$\Omega^2(T,\mu)=\Lambda_\mathrm{3M}^2+m^2(T,\mu)$.  Hence we reproduce the
phenomenology of the the NJL model, in particular the gap-equation and the
chiral phase transition. The NJL model constitutes a nontrivial example of the
equivalence between instant form and front form quantization. We argue that
the light front formulation of thermal field theory is a very useful and
powerful alternative tool to make closer connection to QCD, in particular for
the light front formulation of QCD, e.g., along the lines
of~\cite{Brodsky:1997de}.  The region of interest is that of finite $\mu$,
where lattice QCD just begins to become available for small
$\mu$~\cite{Fodor:2002sd}. Further merits of the light front approach are
obvious when considering multi-quark correlations as has been already outlined
in~\cite{Beyer:2001bc,Mattiello:2001vq}.

\section{Acknowledgment}
Work supported by Deutsche Forschungsgemeinschaft, grant BE 1092/10. TF thanks
to Funda\c c\~ao de Amparo a Pesquisa do Estado de S\~ao Paulo (FAPESP) and to
Conselho Nacional de Desenvolvimento Cient\'\i fico e Tecnol\'ogico of Brazil
(CNPq).

\begin{appendix}
\section{Light-front notation}\label{app:notation}
The contravariant components of a four
vector in  light-front coordinates may be written as
\begin{equation}
a^\mu=(a^-,a^+,a^1,a^2)\equiv(a^-,a^+,\vec a_\perp)\equiv(a^-,\vk{a})
\end{equation}
If we define $a^\pm\equiv a^0\pm a^3$, the standard {\sc Lorentz} scalar
product, can be recovered by defining the following metric tensor
\begin{equation}
g_{\mu\nu}=
\left(
\begin{array}{cccc}
{0}&{\frac{1}{2}}&{0}&{0}\\
{\frac{1}{2}}&{0}&{0}&{0}\\
{0}&{0}&{-1}&{0}\\
{0}&{0}&{0}&{-1}\end{array}
\right),
\; g^{\mu\nu}=
\left(
\begin{array}{cccc}
{0}&{2}&{0}&{0}\\
{2}&{0}&{0}&{0}\\
{0}&{0}&{-1}&{0}\\
{0}&{0}&{0}&{-1}\end{array}
\right),
\end{equation}
i.e.
\begin{equation}
a_\nu b^\nu=g_{\mu\nu} a^\mu b^\nu = \frac{1}{2}(a^-b^++a^+b^-)-
{\vec a}_\perp{\vec b}_\perp.
\end{equation}
Hence the covariant components of the four vector are given by
\begin{equation}
a_\mu=(a_-,a_+,a_1,a_2)=
({\textstyle \frac{1}{2}a^+,\frac{1}{2}a^-},-{\vec a}_\perp)
\equiv (a_-,\uk{a}).
\end{equation}
Note
\begin{equation}
a_-=\frac{1}{2}a^+, \quad 
a_+=\frac{1}{2}a^-.
\end{equation}
The four-dimensional volume element is
\begin{equation}
d^4x=dx^0dx^1dx^2dx^3=\frac{1}{2}dx^-dx^+d^2x_\perp
\end{equation}
The three-dimensional volume element on the light-like surface $S_\nu$
characterized by the light-like vector  $\omega_\nu$ in the plane with
$\omega_\nu x^\nu = x^+=\mathrm{const}$, is
\begin{equation}
d^4x\big|_{S_+}=\frac{1}{2}dx^-d^2x_\perp\equiv dS_+.
\label{eqn:dS}
\end{equation}
The scalar product of the light-like surface element $dS_\nu$ with a vector
$a^\nu$ is then
\begin{equation}
dS_\nu a^\nu =\frac{1}{2}dx^-d^2x_\perp\; a^+.
\end{equation}
%% For the k-volume we use
%% \begin{equation}
%%   d\Gamma_k\equiv\frac{dk^+d^2\vec k_\perp}{2k^+(2\pi)^3} \theta(k^+)
%% =\frac{dxd^2\vec k_\perp}{2x(2\pi)^3} \theta(x)
%% \label{eqn:gamma}
%% \end{equation}
From the $\gamma$-matrices $\gc^\mu=(\gc^+,\gc^-,\vec \gc_\perp)$ it is
possible to define Hermitian projection operators
\begin{equation}
  \label{eqn:lambda}
\Lambda^\pm=\frac{1}{4}=\gamma^\mp\gamma^\pm.  
\end{equation}
\end{appendix}

\end{document}